\documentclass[runningheads]{llncs}

\usepackage[T1]{fontenc}
\usepackage{graphicx}
\usepackage{hyperref}
\usepackage{cite}
\usepackage{subcaption}
\usepackage{booktabs}
\usepackage[table]{xcolor}
\usepackage{amsmath}

\usepackage{listings}
\usepackage{tikz}
\usepackage{pgfplots}
\pgfplotsset{compat=1.18}

\usetikzlibrary{patterns}

\usepackage{float}
\newfloat{listing}{tbp}{lol}
\floatname{listing}{Listing}

\usetikzlibrary{shapes.geometric, arrows, calc, positioning}

\newcommand{\CC}{C\nolinebreak\hspace{-.05em}\raisebox{.4ex}{\tiny\bf +}\nolinebreak\hspace{-.10em}\raisebox{.4ex}{\tiny\bf +}}
\def\CC{{C\nolinebreak[4]\hspace{-.05em}\raisebox{.4ex}{\tiny\bf ++}}}

\newcommand{\sycldb}{\textbf{SYCLDB}}

\DeclareCaptionSubType{listing}

\begin{document}

\title{SYCLDB: A Cross-Vendor Heterogeneous Analytical DBMS}
\titlerunning{SYCLDB: A Cross-Vendor Heterogeneous Analytical DBMS}

\author{Ivan Donchev Kabadzhov \and Eugenio Marinelli \and Raja Appuswamy}

\institute{EURECOM, Biot, France}
\maketitle

\begin{abstract}

The rapid growth of AI and analytical workloads has increased the need for fast, interactive analytical data processing. This, in turn, has led to a surge of interest in designing accelerated analytical database engines that can exploit modern GPUs to scale computation. While GPUs provide the massive data-parallelism required by modern analytics, the software landscape remains dominated by closed-source, vendor-locked programming models, like CUDA, limiting portability across multi-vendor GPUs. In this work, we present SYCLDB, an end-to-end analytical database engine built upon the SYCL open standard, designed to enable efficient, hardware-agnostic execution across a wide range of accelerators using a single codebase. By implementing core relational operators as modular SYCL kernels, integrating them with a SQL frontend based on Apache Calcite and Apache Thrift, and supporting Apache Arrow for efficient columnar ingestion, SYCLDB reduces dependence on vendor-specific GPU programming stacks by enabling a single SYCL codebase to execute analytical query pipelines across Intel, AMD, and NVIDIA GPUs.


\end{abstract}

\section{Introduction}

The growing popularity of AI and analytics in virtually all application domains has led to a rapid increase in the amount of data gathered by enterprises and scientific institutions alike, with the global datasphere being projected to reach 250ZB by the year 2025. Modern analytics and AI workloads are incredibly diverse and consist of a range of scalar, vector, and matrix computations. Thus, the traditional workhorse of the computing industry, the general-purpose CPU, cannot be optimized to meet the diverse requirements of such heterogeneous workloads. This shortcoming of CPUs, in combination with the cessation of Dennard scaling, has led to a growing interest in the adoption of GPU-accelerated database engines that can meet the demands of large-scale data analytics workloads. Unfortunately, most popular solutions today are developed to target primarily NVIDIA GPUs, and hence build on the proprietary, closed-source CUDA software stack that hinders their portability to GPUs and accelerators from other vendors. Thus, there is a need for developing open, standards-based, cross-vendor approach to hardware acceleration to democratize data analytics.

Over the past few years, SYCL has emerged as a higher-level programming model that brings hardware acceleration to mainstream \CC~with the goal of making it easy to write new software similar to CUDA. Expressing data-parallel computations in SYCL provides two key advantages over other competing programming models. First, SYCL builds on a portable high-level programming language by implementing an open standard in contrast to other data-parallel languages that are proprietary (CUDA) or low level in nature (OpenCL). In doing so, SYCL enables the development of portable, cross-architecture applications. Second, SYCL is a single-source by nature as it allows host and device code to co-exist in a single file unlike other programming models. This allows the SYCL compiler the ability to analyze and optimize across the entire program regardless of the device on which the code is to be run. 

Recent research has investigated the performance--portability trade offs in (i) using SYCL for accelerating key HPC applications~\cite{sycl-hpc-iwocl}, and (ii) utilizing isolated aspects of the SYCL programming model~\cite{9654235, sycl-bench, Lal2020SYCLBenchAV}. Only one prior work (XJoin~\cite{xjoin-eugenio}) focused on using SYCL in the context of database engine design, but it was limited to investigating kernel-level performance benchmarking of isolated projection and hash join operators. Thus, despite this work, there are several open questions that have not been well studied. First, kernel-level microbenchmarks show only one aspect of overall performance. To understand the utility of SYCL for database engine design, it is necessary to implement a full-fledged SYCL-based query executor that can interpret complex query plans from application benchmarks and compare the SYCL-based design with state-of-the-art CUDA counterparts~\cite{crystal,heavydb,Sirius}. Second, to understand the ability to use SYCL for developing next-generation analytical engines, it is also necessary to investigate the ability to integrate SYCL toolchains with pluggable SQL front-ends to demonstrate fully-interpreted, end-to-end SQL query execution.

In this work, we present \textbf{SYCLDB}--a cross-architecture, performance-portable, query execution engine that can accelerate SQL query execution using not just NVIDIA GPUs (like all CUDA-based related work~\cite{Yogatama2022,Yogatama2024,Sirius,Cao2023,heavydb}), but also GPUs from other vendors (like Intel Flex, AMD Instinct, etc.) using a single code base. At its core, \textbf{SYCLDB} implements key relational operators as a sequence of highly tuned, independent kernels in SYCL. To drive SYCL kernels from SQL frontend, \textbf{SYCLDB} integrates Apache Calcite~\cite{calcite} and Apache Thrift~\cite{apache_thrift} to build a composable query parsing/optimization front end that can transform SQL queries into a directed acyclic graph of SYCL kernels.

Using \textbf{SYCLDB}, we perform a rigorous multi-platform evaluation on NVIDIA, AMD, and Intel hardware, demonstrating that: (i) in end-to-end query execution, \textbf{SYCLDB} outperforms state-of-the-art GPU database systems (achieving up to $1.22\times$ speedup on average over HeavyDB and up to $4.72\times$ over SiriusDB) as well as the leading in-memory CPU database DuckDB by $4.7\times$ to $10.7\times$ ($6.37\times$ on average); and (ii) \textbf{SYCLDB} is fully portable across multi-vendor GPUs, enabling end-to-end SQL query execution on all evaluated platforms and demonstrating both the high performance portability of SYCL and its readiness to integrate with standard SQL front-ends. We make \sycldb{} publicly available\footnote{https://gitlab.eurecom.fr/dislab/sycldb} for further research on standards-based analytics acceleration.
\section{Design}
\label{sec:design}

In this section, we present the design of \textbf{SYCLDB} library while highlighting key aspects of SYCL relevant to the design of \textbf{SYCLDB}. 

\subsection{Memory Management}

\sycldb{} is optimized for high-throughput analytical workloads. Thus, similar to state-of-the-art GPU database engines~\cite{crystal,heavydb,Sirius}, it adopts a columnar storage architecture where each table attribute is maintained as a separate, contiguous array. \sycldb{} supports data stored in Apache Arrow-compatible columnar buffers to enable direct data ingestion from external sources. \sycldb{} is also presently designed to work only in the scenario where all data is resident in GPU memory similar to other GPU database engines. As our goal is to be portable across multi-vendor GPUs, we do not assume the existence of high-performance interconnect like NVLink, but rather only low-bandwidth PCIe links connecting the host CPU and GPU. In such scenarios, it is well known~\cite{crystal} that PCIe data transfers become a bottleneck in query execution. To avoid this, similar to other GPU engines, \sycldb{} reads all input columns during the startup phase and moves them to GPU memory. CUDA-based database engines rely on explicit functions for allocating and transferring memory between host (CPU) and device (GPU) memory, such as \texttt{cudaMalloc} and \texttt{cudaMemcpy}. Additionally, CUDA offers pinned memory, which ensures that data transfers between host and device are more efficient. In SYCL, similar functionalities are available via the Unified Shared Memory (USM) model~\cite{sycl2020} which we use to migrate data from host memory to GPU device memory.

\subsection{SYCLDB Primitives}

At its core, \textbf{SYCLDB} is organized as a library of relational primitives that are used as building blocks for implementing higher-level relational operators. There are four fundamental primitives, namely, selection, projection, build, and probe.

\subsubsection{Selection and Projection}

In order to support selection and projection relational operations, \textbf{SYCLDB} supports a generic per-element selection primitive that evaluates a boolean predicate. SYCLDB currently targets fixed-width numeric columns, where selection predicates are expressed as binary comparisons between an input column and a constant reference value. To support multiple comparison operations, we implement selections as basic switch statements of functionals, deciding the binary operation. Figure~\ref{lst:selection_a} shows a basic selection primitive that operates on a single element (\texttt{col}). It compares \texttt{col} with the value \texttt{v} using the comparison type \texttt{comp} and sets the resulting boolean value.

\begin{listing}[t!]
  \centering

  \begin{sublisting}{\textwidth}
    \begin{lstlisting}[basicstyle=\ttfamily\small]
    template <typename T, typename Compare>
    bool selection_primitive(const T& col, const T& v) {
        Compare comp{};
        return comp(col, v);
    }
    \end{lstlisting}
    \vspace{-7mm}
    \caption{Generic selection primitive acting on one element}
    \label{lst:selection_a}
  \end{sublisting}

  \begin{sublisting}{\textwidth}
    \begin{lstlisting}[basicstyle=\ttfamily\small]
    
    // Pass 1: First selection predicate
    bool *flags = sycl::malloc_device<bool>(F.len, q);

    q.parallel_for(sycl::range<1>(F.len), [=](sycl::id<1> idx) {
        size_t i = idx[0];
        flags[i] = selection_primitive<int, sycl::less<>>(F.data_A[i], 100);
    });
    \end{lstlisting}
    \vspace{-7mm}
    \caption{Invoking a selection primitive from a SYCL kernel}
    \label{lst:sycl-one-kernel}
  \end{sublisting}

  \begin{sublisting}{\textwidth}
    \begin{lstlisting}[basicstyle=\ttfamily\small]

    
    // Pass 1: First selection predicate
    bool *flags = sycl::malloc_device<bool>(F.len, q);

    q.parallel_for(sycl::range<1>(F.len), [=](sycl::id<1> idx) {
        size_t i = idx[0];
        flags[i] = selection_primitive<int, sycl::less<>>(F.data_A[i], 100);
    });

    // Pass 2: Second selection predicate
    q.parallel_for(sycl::range<1>(F.len), [=](sycl::id<1> idx) {
        size_t i = idx[0];
        flags[i] &= selection_primitive<int, sycl::less<>>(F.data_B[i], 50);
    });
    \end{lstlisting}
    \vspace{-7mm}
    \caption{Kernel-at-a-time execution strategy}
    \label{lst:sycl-multiple-kernels}
  \end{sublisting}

  \caption{SYCLDB primitive and operator execution strategy}
  \label{lst:selection_comparison}
\end{listing}

\subsubsection{Build and Probe}

To support the relational hash join operator, \sycldb{} supports two core primitives: \textit{build} and \textit{probe}. The current implementation targets hash-based equi-joins over fixed-width key columns, matching the join patterns commonly found in star-schema analytical workloads such as the Star Schema Benchmark (SSB)~\cite{ssb}. We utilize a pointer-free, array-based hash table with linear probing and simple remainder hashing. The \textit{build} primitive populates the hash table from the build relation. The \textit{probe} primitive then evaluates matches from the probe relation.

\subsection{Relational Operators}

\textbf{SYCLDB} uses the individual primitives listed earlier to implement higher-level relational operators that process one or more input relations in a massively parallel fashion across different GPU backends. Similar to other accelerator programming approaches, the SYCL model makes the distinction between a host CPU and an accelerator device. The key data-parallel programming constructs in SYCL are \emph{kernels}, whose execution is coordinated from the host. SYCL provides the abstraction of a \emph{queue} to enable host code to command devices. Queues accept work requests as submissions, where each work request describes a data-parallel kernel. Submission is an asynchronous operation, and control returns back to the host code after the request has been queued for execution on the associated device. 
Unlike CUDA which requires kernel dimensions to be explicitly specified, SYCL allows developers to specify a simple 1D-range that spans a given input array that needs to be processed, with the compiler determining the most effective data partitioning strategy across GPU threads. \textbf{SYCLDB} uses this functionality to simplify the implementation of relational operator kernels. 

\subsubsection{Selection operator.} Each selection condition has to apply a binary operation between a column and a fixed reference value (that is \texttt{SELECT $x$ FROM $R$ WHERE $x < v$}). Multiple such conditions can appear in the same query. To enable efficient, materialization-free condition evaluation, \sycldb{} keeps track of selection flag vectors for each relation similar to other analytical engines. These vectors encode whether each tuple remains active after the predicates evaluated so far. For conjunctive predicates, each new predicate result is combined with the previous flag value, so a tuple remains active only if it satisfies all conditions processed up to that point.
The code in Listing~\ref{lst:selection_comparison} shows the code corresponding to a selection operator with two conditions. The code contains several key variables: a queue (\texttt{q}) representing the device (GPU) context, the input relation (\texttt{F}) to be processed, a selection flags array (\texttt{sf}) to store the results of the comparison operation, and the cardinality of the relation (\texttt{F.len}). Additionally, it includes comparison operator (\texttt{sycl::less}), as well as a constant value (\texttt{100}/\texttt{50} for first/second conditions) against which the column (\texttt{data\_A}/\texttt{data\_B}'s entries for first/second conditions) are compared. 
In Listing~\ref{lst:selection_comparison}, the selection primitive evaluates a single predicate and returns its boolean result. Predicate composition is handled by the operator kernel. The first pass initializes the flag vector, while subsequent passes update each flag by combining the previous value with the newly evaluated predicate. The kernel's body employs a simple parallel-for loop over a 1D-range of size \texttt{F.len}. Inside the loop, the selection primitive is called with the appropriate arguments for each element at index \texttt{i}. 
This design keeps each predicate pass compact and allows the frontend to compose arbitrary conjunctions of supported predicates while preserving a portable kernel-at-a-time execution model. At present, \sycldb{} supports fixed-width numeric attributes and simple comparison predicates, such as $=$, $<$, $>$, $\leq$, and $\geq$, which are executed directly inside SYCL kernels.

\subsubsection{Join operator.} The join operator is implemented by two separate kernel invocations, each mapping the build and probe primitives respectively. The current operator supports equi-joins over fixed-width keys, which are sufficient to express the Star Schema Benchmark (SSB) join patterns used in our evaluation. The hash table produced by the build primitive is materialized and made available to the probe kernel via GPU global memory. The probe kernel also materializes its result vector of joined values in GPU global memory for consumption by downstream operators (like aggregation).

\subsubsection{Aggregation operator.} 
\begin{figure}[t!]
  \centering
\begin{tikzpicture}[
    every node/.style={font=\footnotesize},
    array/.style={rectangle, draw, minimum size=0.5cm, text centered, font=\footnotesize},
    global/.style={rectangle, draw, fill=gray!10, minimum size=0.35cm},
    arrow/.style={->, thick},
    dashedarrow/.style={->, thick, dashed, gray}
]

\node at (-1.5, 1.5) {\bf Global Data};
\foreach \x in {0,...,15} {
    \node[global] at (\x*0.45, 1.5) (g\x) {};
}

\node at (-1.5, 0) {Step 0: Streaming};
\foreach \x in {0,...,3} {
    \node[array, fill=green!5] at (\x*1.8 + 0.45, 0) (wi\x) {Acc};
}

\foreach \x in {0,1,2,3} {
    \draw[dashedarrow] (g\x.south) -- (wi\x.north);
    \pgfmathtruncatemacro{\next}{\x+4}
    \draw[dashedarrow] (g\next.south) -- (wi\x.north);
    \pgfmathtruncatemacro{\last}{\x+8}
    \draw[dashedarrow] (g\last.south) -- (wi\x.north);
}

\node at (-1.5, -1.1) {Step 1: Tree (L/2)};
\foreach \x in {0,...,3} {
    \node[array] at (\x*1.8 + 0.45, -1.1) (t1c\x) {\ifnum\x<2 $\sum$\else\fi};
}

\node at (-1.5, -2.2) {Step 2: Tree (L/4)};
\foreach \x in {0,...,3} {
    \node[array] at (\x*1.8 + 0.45, -2.2) (t2c\x) {\ifnum\x<1 $\sum$\else\fi};
}

\draw[arrow] (wi2.south) -- (t1c0.north);
\draw[arrow] (wi3.south) -- (t1c1.north);
\draw[arrow] (wi0.south) -- (t1c0.north);
\draw[arrow] (wi1.south) -- (t1c1.north);

\draw[arrow] (t1c1.south) -- (t2c0.north);
\draw[arrow] (t1c0.south) -- (t2c0.north);


\end{tikzpicture}
  \caption{SYCL2020 Reduction: Register Streaming + Shared-Memory Tree}
  \label{sycl_reduce}
\end{figure}
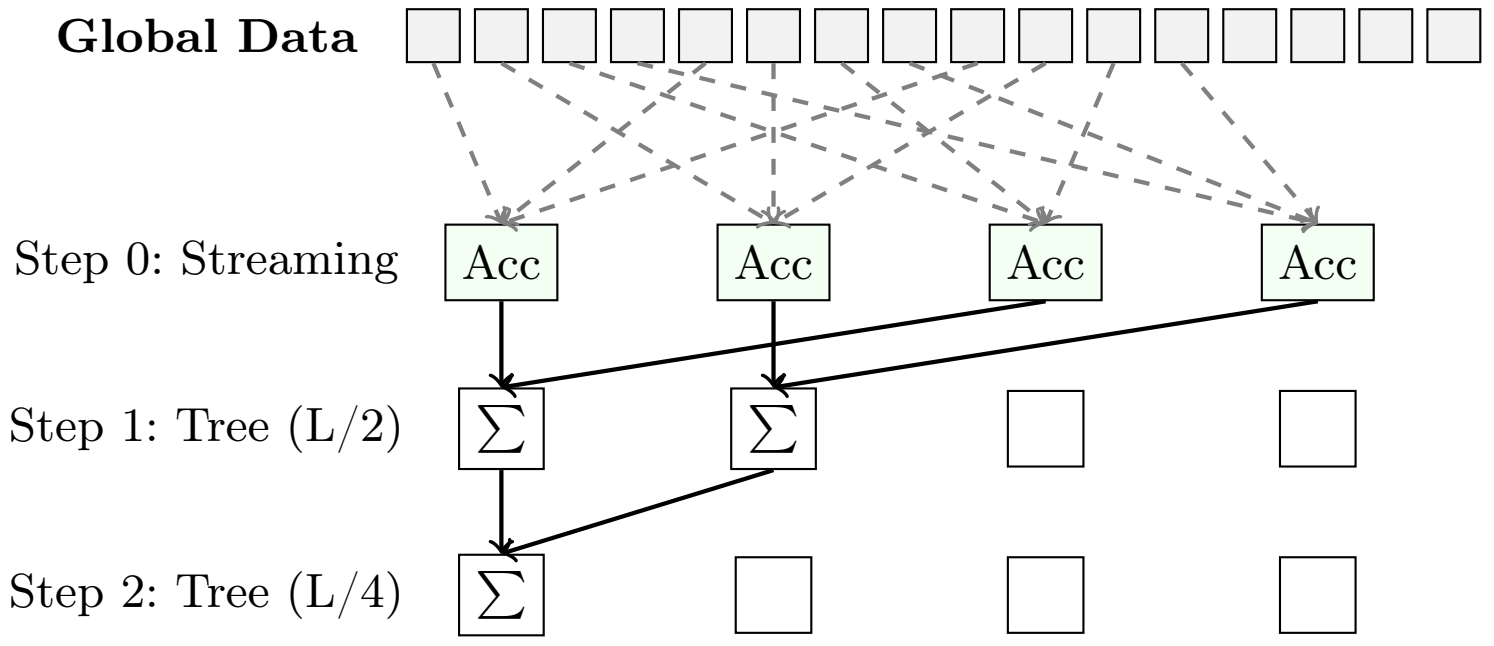

Global aggregation is essential for executing SQL operators such as \texttt{SUM}, \texttt{COUNT}, and \texttt{MIN/MAX}. SYCLDB currently supports these aggregations over fixed-width numeric columns, where the aggregation operation is associative and commutative. State-of-the-art GPU implementations of global aggregation, like Crystal~\cite{crystal}, often explicitly orchestrate a multi-level hierarchy: using sub-group shuffles for intra-warp communication, local memory for inter-warp synchronization, and global atomics for inter-group finalization. For instance, Figure~\ref{sycl_reduce} illustrates a \texttt{SUM} reduction. Each work-item in step zero loops internally to stream data from global memory into a local register accumulator (\texttt{Acc}). This strategy ensures perfectly coalesced memory accesses and hides instruction latency through deep pipelining. Once the streaming phase is complete, the runtime automatically transitions to Step 1 and Step 2, performing an efficient tree-based reduction within \textit{local memory}. As a result, the final inter-group reduction involves a minimal number of partial results, which drastically reduces contention on global atomic units. 
While this approach provides high performance, it has the limitation that it is hardware specific and requires complex fine tuning corresponding to the underlying device.
With SYCL, we can express this entire high-performance pipeline in a single line of code that leverages a key enhancement introduced in the SYCL 2020 revision: the \emph{reducer} class--equipped with built-in support for common aggregation functions, such as sum, product, minimum, and maximum, making it highly versatile. Listing~\ref{lst:reduce_comp} shows an example reduction that performs an entire associative-commutative reduction over \texttt{data[0..N)} into the scalar location \texttt{*sum} in a single kernel call, leaving the choice of in-kernel reduction strategy (work-group tree reduction, subgroup shuffle, final atomic merge) to the SYCL runtime.

Reading the call left to right, we can see that the code submits a single, 1-D range kernel to the queue \texttt{q}. The second argument, \texttt{sycl::reduction(sum, sycl::plus<>())}, is a \emph{reduction descriptor} that declares that there is a
scalar accumulator at the address \texttt{sum}, to be combined using \texttt{sycl::plus<>()}. This is followed by a lambda whose parameter \texttt{sum} deliberately shadows the outer pointer of the same name. Inside the kernel, \texttt{sum} is no longer a pointer to memory but a  \emph{reducer} object supplied by the runtime. The reducer accumulates thread-local (or \texttt{work-item}-local in SYCL) partials, the runtime combines partials across the threadblock (or \texttt{work-group} in SYCL) using whatever shared-memory primitive the target backend exposes, and a small number of threadblocks perform the final atomic update against the underlying location.
This single line of code is semantically equivalent to the multi-stage aggregation pipeline implemented by optimized systems like Crystal and is tuned by the SYCL runtime automatically for each target backend.

\subsubsection{GROUP BY operator.} State-of-the-art GPU implementations of the \texttt{GROUP BY} operation would either sort the input by group key and run a segmented reduction, or build per-block partial histograms in shared memory and then merge them globally.  Both alternatives trade off more memory traffic and more kernel launches for reduced contention with atomics. More recently, based on the fact that modern GPUs are much better at handling warp and block-level atomics, Crystal~\cite{crystal} proposed a tiled implementation. Each
thread block builds a \emph{private} hash table in shared memory,
aggregates its own tile against that local table using atomics (\texttt{atomicCAS} and \texttt{atomicAdd} on \texttt{\_\_shared\_\_} locations), synchronizes once with \texttt{\_\_syncthreads()}, and then merges the local table into the global result with one
\texttt{atomicAdd} per occupied slot. This implementation assumes that the number of groups in a \texttt{GROUP BY} is  small enough to fit in shared memory. When this is not the case,  Crystal falls back to direct global atomics with \texttt{atomicAdd}. The downside of this approach is that it is hardware specific, as it relies on three different CUDA synchronization mechanisms--warp shuffles and shared-memory atomics internally, a block-wide barrier between phases, and the fact that NVCC compiler lowers the \texttt{atomicAdd} it to a PTX \texttt{atom.relaxed.gpu.add.u64} instruction, which provides relaxed memory ordering semantics. Relaxed memory ordering is safe and valuable here precisely because of the algebraic structure of \texttt{GROUP BY} with a summing aggregate. Each row contributes to exactly one bucket. As buckets are disjoint, there is no cross-bucket dependence that another thread might need to observe in order. Within a bucket, the only operation is \texttt{+=}, which is both associative and commutative, so any interleaving of \texttt{fetch\_add}s from different rows produces the same result. 

\textbf{SYCLDB} trades Crystal's locality optimization for code uniformity and backend portability. Instead of two different implementations that depend on the group size, SYCLDB in contrast, uses the \texttt{sycl::atomic\_ref} SYCL feature to achieve the same effect in a portable fashion. An \texttt{atomic\_ref<T, Order, Scope, Space>} is a \emph{non-owning} reference that re-interprets reads, writes, and read-modify-writes
through it as atomic operations on the underlying storage. A key template parameter taken by \texttt{atomic\_ref} is the choice of memory model.
The use of \texttt{memory\_order::relaxed} results in the SYCL compiler producing the optimal code for each GPU backend, namely the same PTX instruction as CUDA \texttt{atomicAdd} on NVIDIA GPUs, a global atomic without an explicit \texttt{s\_waitcnt vmcnt(0)} on AMD CDNA devices, and a Level-Zero atomic without cache-line invalidation on Intel Xe GPUs.

\textbf{SYCLDB} uses \texttt{atomic\_ref} to perform grouped aggregation in a single \texttt{parallel\_for} over the input column with no intra-block reduction and no explicit synchronization. In order to do this, it assumes that the workload has a Star schema similar to other GPU engines where each dimension column has a small, contiguous range of integer codes. Hence, the Cartesian product of any handful of group-by columns fits comfortably in a flat array, and the hash function collapses to mixed-radix arithmetic with no collisions. 
Building on the dense-key assumption introduced above, each active row computes its slot in the perfect-hash result array using mixed-radix arithmetic over the group-key domains. 
Listing~\ref{lst:sycldb-groupby} shows the GROUP BY implementation. After computing the slot, each row issues two relaxed device-scope atomics: an \texttt{exchange} that elects the first writer responsible for installing the representative group keys, and a \texttt{fetch\_add} that accumulates the numeric aggregate.

\begin{listing}[t!]
    \begin{lstlisting}[basicstyle=\ttfamily\footnotesize]
q.parallel_for(N, [=](sycl::id<1> idx) {
    const auto i = idx[0];
    if (!flags[i]) return;

    int hash = 0, mult = 1;
    for (int j = 0; j < G; ++j) {
        hash += (group_cols[j][i] - min[j]) * mult;
        mult *= max[j] - min[j] + 1;
    }

    sycl::atomic_ref<unsigned, sycl::memory_order::relaxed,
        sycl::memory_scope::device,
        sycl::access::address_space::global_space> first(slot_flags[hash]);
    if (first.exchange(1) == 0)
        for (int j = 0; j < G; ++j) keys[j][hash] = group_cols[j][i];

    sycl::atomic_ref<uint64_t, sycl::memory_order::relaxed,
        sycl::memory_scope::device,
        sycl::access::address_space::global_space> sum(agg[hash]);
    sum.fetch_add(value[i]);
});
\end{lstlisting}
\caption{SYCLDB GROUP BY implementation.}
\label{lst:sycldb-groupby}
\end{listing}

\begin{listing}[t!]
\begin{lstlisting}[basicstyle=\ttfamily\footnotesize]
q.parallel_for(range, sycl::reduction(sum, sycl::plus<>()),
  [=](sycl::id<1> id, auto& sum) { sum += data[id]; });
\end{lstlisting}
\caption{Modular aggregation using the high-level SYCL 2020 reduction API}
\label{lst:reduce_comp} 
\end{listing}

\subsection{Interpreted Query Execution}

To execute SQL queries dynamically on modern GPU accelerators, \textbf{SYCLDB} employs a decoupled server-client query planning and execution architecture. A Java Virtual Machine based optimizer using Apache Calcite~\cite{calcite} parses and optimizes the incoming SQL query, applying cost-based relational rules to perform join and operator reordering. The resulting optimized physical execution plan is serialized and transmitted via Apache Thrift RPC to the \CC{} execution engine, which deserializes this plan and constructs a directed acyclic graph (DAG) representing the physical operator execution order.
SYCLDB supports the following physical operators: scans over columnar inputs, selections, projections, equi-joins, global aggregations, and grouped aggregations. More general SQL features, including nested queries, user-defined functions, full NULL semantics, arbitrary string predicates, and general-purpose grouping over sparse keys, are outside the current prototype.
\textbf{SYCLDB} executes the DAG by dynamically traversing it in topological order and dispatching pre-compiled, template-based general-purpose operators directly to the device queue. Thus, \sycldb{} performs kernel-at-a-time execution to evaluate any given SQL query. It is important to note that while we do kernel-at-a-time execution, we do not perform full materialization after each kernel due to use of selection flags. Instead, materialization happens at natural pipeline boundaries similar to other vectorized database engines. As modern SYCL compilers have started adding support for efficient JIT-compiled fusion of SYCL kernels, we leave open the task of supporting fused kernel execution in \sycldb{} to future work.

\section{Evaluation}

\textbf{Hardware Platforms.} We evaluate \textbf{SYCLDB} across three heterogeneous GPU architectures, each connected via a PCIe Gen4 x16 interconnect (31.5 GB/s theoretical bi-directional bandwidth). The host system is equipped with a dual-socket \textbf{Intel Xeon Platinum 8454H} 
processors (Sapphire Rapids, 32 cores/socket, 128 threads total, 3.4GHz boost), 
\textbf{1\,TB DDR5 ECC} system memory, and a Gen4 NVMe SSD capable of 
7,400\,MB/s sequential reads, used for initial data staging into system memory.

\begin{itemize}
    \item \textbf{NVIDIA L40S:} Based on the Ada Lovelace architecture, featuring \textbf{48 GB of GDDR6} memory with a peak bandwidth of \textbf{864 GB/s}. It operates at a 350W TDP and provides high-throughput FP32/INT32 compute units.
    \item \textbf{AMD Instinct MI210:} Built on the CDNA 2 architecture, providing \textbf{64 GB of HBM2e} memory with a peak bandwidth of \textbf{1.6 TB/s}. The HBM2e stack is critical for saturating the high-throughput requirements of analytical workloads at scale.
    \item \textbf{Intel Data Center GPU Flex 170:} Based on the Xe-HPG architecture, featuring \textbf{16 GB of GDDR6} memory with a peak bandwidth of \textbf{576 GB/s}. This platform represents a high-density, power-efficient acceleration point for moderate-scale datasets.
\end{itemize}

\textbf{Dataset and Workload.} We utilize SSB as the main benchmark similar to other recent works on GPU database engines. We vary the Scale Factor (SF) depending on the backend GPU used in each experimental setting to accommodate memory capacity restrictions of the Intel GPU. The exact SF is mentioned with each experiment. To isolate the engine's compute and memory hierarchy performance, all queries are executed after a full columnar load into VRAM, ensuring the working set is device-resident.

\textbf{Software Stack.} We use \textbf{AdaptiveCpp 25.10.0} (compiled with LLVM 18) as our primary SYCL toolchain for NVIDIA, AMD, and Intel GPUs, targeting the CUDA backend (CUDA 12.6 SDK), HIP backend (ROCm 6.3.3), and OpenCL backend (Intel Compute Runtime), respectively.

\textbf{Measurement Protocol and Baseline Normalization.} To eliminate transient hardware noise and cold-start compilation overhead, each query is first executed as an unrecorded cold pass. Reported timings correspond to the steady-state performance averaged over 5 subsequent consecutive runs. Because the steady-state variance across execution loops was under 10\%, error bars are omitted from results. Unless otherwise stated, query execution time excludes initial table loading from persistent storage and assumes that input columns are already resident in the memory space used by each engine. For GPU engines, kernel-only time includes only device kernel execution, while total time includes SQL planning where applicable, host-side operator orchestration, kernel launch overheads, synchronization points, and device-side intermediate materialization. For systems that perform JIT compilation, compilation overhead is excluded from kernel-only time and included in total time only when it occurs as part of the measured query execution path.

\subsection{End-to-End Query Execution Performance}

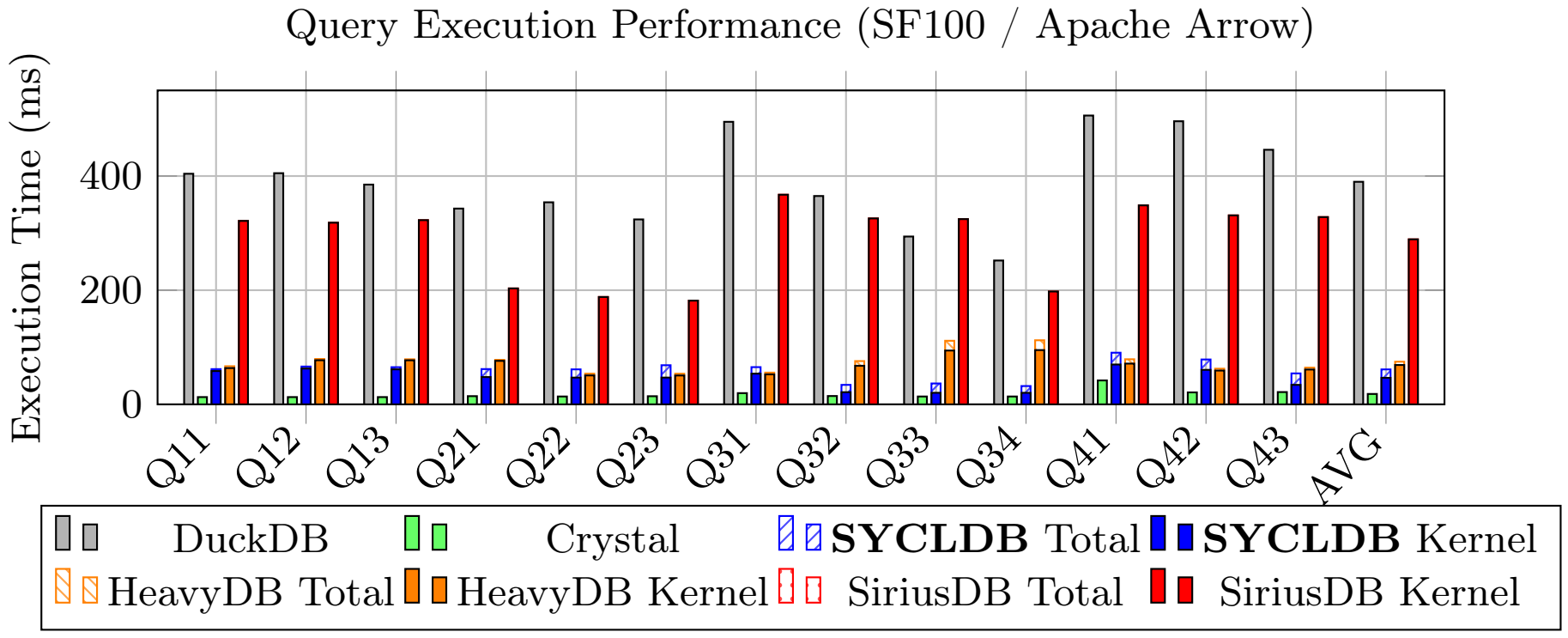
\begin{figure}[t!]
\centering
\begin{tikzpicture}
\begin{axis}[
    ybar,
    width=11.5cm,
    height=4cm,
    bar width=2.0pt,
    ylabel={Execution Time (ms)},
    xlabel={SSB Query},
    symbolic x coords={Q11, Q12, Q13, Q21, Q22, Q23, Q31, Q32, Q33, Q34, Q41, Q42, Q43, AVG},
    xtick=data,
    x tick label style={rotate=45,anchor=east},
    legend style={at={(0.5,-0.32)}, anchor=north, legend columns=4, font=\small},
    grid=major,
    ymin=0, ymax=550,
    enlarge x limits=0.05,
    title={Query Execution Performance (SF100 / Apache Arrow)},
]

\addplot[bar shift=-6pt, fill=gray!60, draw=black]
coordinates {
(Q11,404.0) (Q12,405.0) (Q13,385.0)
(Q21,343.0) (Q22,354.0) (Q23,324.0)
(Q31,495.0) (Q32,365.0) (Q33,294.0) (Q34,252.0)
(Q41,506.0) (Q42,496.0) (Q43,446.0) (AVG,389.92)
};

\addplot[bar shift=-3pt, fill=green!60, draw=black]
coordinates {
(Q11,12.73) (Q12,12.73) (Q13,12.74)
(Q21,14.37) (Q22,13.77) (Q23,14.25)
(Q31,19.66) (Q32,14.60) (Q33,13.76) (Q34,13.77)
(Q41,41.91) (Q42,20.95) (Q43,21.52) (AVG,18.21)
};

\addplot[bar shift=0pt, pattern=north east lines, pattern color=blue!70, draw=blue]
coordinates {
(Q11,61.73) (Q12,66.23) (Q13,65.12)
(Q21,61.80) (Q22,61.29) (Q23,68.55)
(Q31,65.21) (Q32,34.14) (Q33,36.59) (Q34,32.07)
(Q41,90.48) (Q42,78.59) (Q43,54.22) (AVG,61.23)
};

\addplot[bar shift=0pt, fill=blue, draw=black]
coordinates {
(Q11,58.36) (Q12,62.77) (Q13,61.32)
(Q21,47.84) (Q22,46.88) (Q23,46.92)
(Q31,53.79) (Q32,21.31) (Q33,20.25) (Q34,20.21)
(Q41,69.82) (Q42,60.45) (Q43,34.14) (AVG,46.47)
};

\addplot[bar shift=3pt, pattern=north west lines, pattern color=orange!70, draw=orange]
coordinates {
(Q11,66.80) (Q12,79.50) (Q13,79.00)
(Q21,78.00) (Q22,53.80) (Q23,53.70)
(Q31,55.60) (Q32,75.80) (Q33,111.40) (Q34,112.30)
(Q41,79.00) (Q42,62.50) (Q43,64.30) (AVG,74.75)
};

\addplot[bar shift=3pt, fill=orange, draw=black]
coordinates {
(Q11,63.60) (Q12,77.30) (Q13,77.20)
(Q21,76.20) (Q22,50.70) (Q23,50.60)
(Q31,52.50) (Q32,67.60) (Q33,94.30) (Q34,95.10)
(Q41,71.40) (Q42,59.20) (Q43,61.00) (AVG,68.98)
};

\addplot[bar shift=6pt, pattern=dots, pattern color=red!70, draw=red]
coordinates {
(Q11,321.38) (Q12,318.22) (Q13,322.72)
(Q21,203.09) (Q22,188.08) (Q23,181.69)
(Q31,367.24) (Q32,325.72) (Q33,324.53) (Q34,197.70)
(Q41,348.61) (Q42,331.06) (Q43,328.06) (AVG,289.08)
};

\addplot[bar shift=6pt, fill=red, draw=black]
coordinates {
(Q11,321.38) (Q12,318.22) (Q13,322.72)
(Q21,203.09) (Q22,188.08) (Q23,181.69)
(Q31,367.24) (Q32,325.72) (Q33,324.53) (Q34,197.70)
(Q41,348.61) (Q42,331.06) (Q43,328.06) (AVG,289.08)
};

\legend{
DuckDB,
Crystal,
\textbf{SYCLDB} Total, \textbf{SYCLDB} Kernel,
HeavyDB Total, HeavyDB Kernel,
SiriusDB Total, SiriusDB Kernel
}

\end{axis}
\end{tikzpicture}
\caption{Query execution performance at \textbf{SF100} scale. 
The solid bars report kernel execution time, assuming that the input data is already resident in device memory. The hatched bars include the additional query frontend overhead (SQL parsing, optimization, kernel launches, etc.).}
\label{fig:end-to-end}
\end{figure}

To evaluate SYCLDB under a large-scale analytical workload, we benchmarked the execution time of SYCLDB, HeavyDB (industry-grade GPU database engine with compiled query execution), SiriusDB (composable analytics engine based on composing dataframe kernels), Crystal (hard-coded library of relational primitives), and DuckDB (state-of-the-art CPU-based analytical engine) under the SSB benchmark. For this analysis, we use Scale Factor 100 (SF100), corresponding to approximately 600 million rows.
The results are shown in Figure~\ref{fig:end-to-end}. The solid bar is the kernel-only time while the hatched bars include additional frontend and host-side kernel orchestration overheads.
To make the comparison fair across all systems, data are read from Apache Arrow-backed columnar storage and materialized in the execution memory space required by each engine before timing begins.

DuckDB provides a strong reference point for CPU query execution as it is a highly-optimized, state-of-the-art, open-source in-memory analytical engine. Compared with DuckDB, \textbf{SYCLDB} achieves substantially lower execution times across the benchmark with a speedup between $4.7\times$ and $10.7\times$. These results show the potential of GPUs for analytical query execution and underscore potential for further improvements, especially given the fact that \sycldb{} performs straight-forward kernel-at-a-time execution without optimizations like pipelining, kernel fusion, or sophisticated indexing. On average, DuckDB requires 389.92\,ms per query, whereas SYCLDB Total requires 61.23\,ms (and 46.47\,ms for Kernel time only), which represents a $6.37\times$ speedup in overall end-to-end execution ($8.39\times$ speedup in kernel-only execution) over the CPU baseline.

Crystal, on the other hand, represents a specialized GPU implementation in which SQL queries are directly implemented as hard-coded CUDA kernels, thus, removing SQL parsing, query planning, frontend processing, dynamic specialization, and most general-purpose operator-boundary costs. While Crystal is not built to be a general-purpose query execution engine, it acts as an absolute lower-bound baseline for specialized GPU execution time. As expected, Crystal achieves the lowest execution times across the benchmark, with \textbf{SYCLDB}'s kernel time between $1.46\times$ and $4.93\times$ higher than Crystal across the entire benchmark. For several complex join queries such as Q3.2, Q3.3, Q3.4, the gap is small with \textbf{SYCLDB} only about $1.46\times$--$1.47\times$ slower than the manual Crystal baseline. These results show that the \sycldb{} preserves generality while providing performance competitive with a highly specialized implementation for several queries. Comparing their averages, the hand-written, highly specialized Crystal baseline achieves a mean query time of 18.21\,ms, while SYCLDB's kernel-only execution averages 46.47\,ms, demonstrating that SYCLDB remains competitive within a $2.55\times$ factor of a specialized, non-general-purpose CUDA GPU baseline while preserving full relational generality.

Comparing \sycldb{} to SiriusDB, we see that SiriusDB is $4\times$ - $16.0\times$ slower than \textbf{SYCLDB}. The gap is especially visible for Q3.2 and Q3.3, where SiriusDB requires 325.72\,ms and 324.53\,ms, while \textbf{SYCLDB} executes the same kernel workloads in 21.31\,ms and 20.25\,ms, corresponding to speedups of $15.3\times$ and $16.0\times$, respectively. This is due to the cost of materializing intermediate states across operator boundaries. In \textbf{SYCLDB}, general-purpose execution propagates intermediate condition outcomes across operators through device-resident flag arrays, with additional write and read from global device memory between successive operators. 
This cost is even more pronounced in SiriusDB, which relies on chaining highly optimized \texttt{libcudf} library functions. Although each individual primitive is optimized, the pipeline is fragmented across external library calls. Consequently, intermediate outputs produced by filters and join probes must be materialized in global device memory before the next library operator can begin, leading to substantially higher execution times across the benchmark. Considering kernel execution time, the comparison shows that \textbf{SYCLDB}'s approach of building on SYCL compiler and runtime toolchain is better than a dataframe-library composition model adopted by SiriusDB, as \sycldb{}'s specialized relational kernels keep execution closer to the physical query plan and avoid the overhead incurred by chaining general-purpose library primitives. In terms of macro-level averages, SiriusDB exhibits a mean execution time of 289.08\,ms, which is $4.72\times$ slower than SYCLDB's end-to-end average of 61.23\,ms and $6.22\times$ slower than SYCLDB's kernel-only average of 46.47\,ms, highlighting the significant VRAM materialization tax of dataframe library chaining.

HeavyDB follows a completely different design strategy by using a custom LLVM backend to generate JIT-compiled kernels.
Comparing \sycldb{} and HeavyDB, we see that \textbf{SYCLDB} remains highly competitive. In raw kernel execution time, \textbf{SYCLDB} outperforms HeavyDB on 11 out of the 13 evaluated SSB queries with a significant speedup for the Q3 query family: for Q3.2, \textbf{SYCLDB} executes the kernel workload in 21.31\,ms compared to 67.60\,ms for HeavyDB ($3.2\times$ faster), while for Q3.3 and Q3.4 it requires 20.25\,ms and 20.21\,ms compared to 94.30\,ms and 95.10\,ms for HeavyDB, corresponding to speedups of $4.7\times$ in both cases. HeavyDB is slightly faster in kernel time for Q3.1 and Q4.2, but the difference is less than 5\%: 52.50\,ms versus 53.79\,ms for Q3.1, and 59.20\,ms versus 60.45\,ms for Q4.2. These results show the potential of SYCL for developing GPU-based analytical engines and open the door for further performance optimizations that can be achieved by using JIT compilation to fuse SYCL kernels in the future. Looking at the overall benchmark averages, HeavyDB requires an average of 74.75\,ms for end-to-end execution (and 68.98\,ms for kernel time), whereas SYCLDB requires 61.23\,ms (and 46.47\,ms for kernel time), showing that SYCLDB achieves a $1.22\times$ speedup in end-to-end execution and a $1.48\times$ speedup in kernel execution over HeavyDB's custom-generated LLVM query pipeline.

Finally, the gap between the hatched bars and the solid bars quantifies the overhead of the frontend and kernel orchestration. In \textbf{SYCLDB}, each relational operator is submitted as a separate SYCL kernel and followed by an explicit synchronization point used to enforce execution order and collect profiling information. This overhead includes kernel dispatch, synchronization, and intermediate state management. Across the benchmark, this overhead ranges from approximately 3.4\,ms to 21.6\,ms, with an average of about 13.2\,ms per query. The smallest overheads appear in Q1.1--Q1.3, where the gap between kernel time and total time is only about 3--4\,ms, while more complex queries such as Q2.3 and Q4.1--Q4.3 incur larger orchestration costs. Even with this overhead, \textbf{SYCLDB} remains substantially faster than DuckDB and SiriusDB across all queries, indicating that the general-purpose execution path and the composable front end both remain practical and efficient for large-scale analytical workloads.

\subsection{Performance Portability and Cross-Vendor Analysis}
\label{subsec:performance_portability}

In this section, we analyze the performance portability of \sycldb{} across diverse hardware architectures, validating that (i) the same codebase executes seamlessly across multiple vendor stacks, and (ii) portable SYCL implementations can compete with or outperform native, highly specialized GPU baselines. We compare the full \textbf{SYCLDB} engine against an evaluation-only \textbf{SYCL-Hardcoded} reference implementation. SYCL-Hardcoded is not a separate execution path exposed by \sycldb{}; rather, it uses the same SYCLDB relational primitives inside hand-written query kernels similar to Crystal to estimate the performance available without a SQL front end.

To demonstrate cross-vendor portability, we execute the full SSB benchmark on three distinct GPU architectures: the NVIDIA L40S, the AMD Instinct MI210, and the Intel Data Center GPU Flex 170. We use Scale Factor 40 (SF40) for these experiments, as it represents the memory limit of the Intel Flex 170 device, and Scale Factor 100 (SF100) for the NVIDIA and AMD platforms. We used the ROCm HIPIFY tool~\footnote{\url{https://github.com/ROCm/HIPIFY}} to convert the CUDA Crystal baseline into its HIP version for AMD evaluation. Notably, the hand-written CUDA baseline (Crystal) and its HIP translation are entirely vendor-locked and cannot run on the Intel Flex 170. This highlights the role of open standards like SYCL in building general-purpose, cross-vendor database engines.

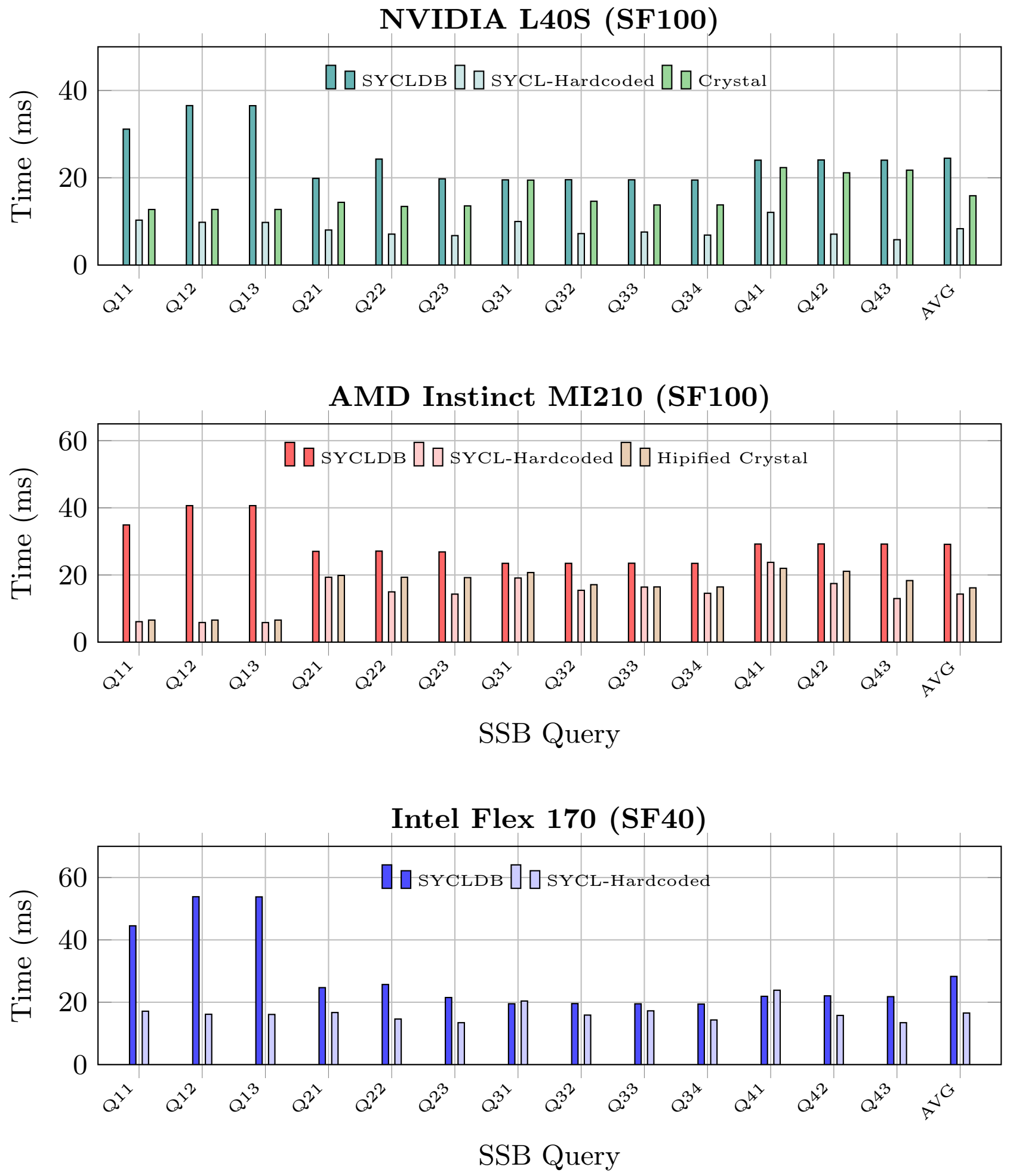
\begin{figure*}[t!]
\centering

\begin{tikzpicture}
\begin{axis}[
    ybar,
    width=0.95\textwidth, 
    height=4.0cm,
    bar width=2.0pt,
    ylabel={Time (ms)},
    symbolic x coords={Q11, Q12, Q13, Q21, Q22, Q23, Q31, Q32, Q33, Q34, Q41, Q42, Q43, AVG},
    xtick=data,
    x tick label style={rotate=45,anchor=east,font=\tiny},
    legend style={at={(0.5,0.95)}, anchor=north, legend columns=3, font=\tiny, draw=none, fill=none},
    grid=major,
    ymin=0, ymax=50,
    enlarge x limits=0.05,
    title={\textbf{NVIDIA L40S (SF100)}},
    title style={font=\footnotesize, yshift=-1.5ex}
]
\addplot[fill=teal!60, draw=black] coordinates {
    (Q11,31.14) (Q12,36.53) (Q13,36.51) (Q21,19.86) (Q22,24.28) (Q23,19.75) 
    (Q31,19.52) (Q32,19.55) (Q33,19.53) (Q34,19.48) (Q41,24.03) (Q42,24.08) (Q43,24.03) (AVG,24.48)
};
\addplot[
    fill=teal!20, draw=black,
    nodes near coords,
    every node near coord/.style={font=\tiny, color=black, yshift=-2pt, anchor=south},
    point meta=explicit symbolic
] coordinates {
    (Q11,10.29) (Q12,9.80) (Q13,9.77) 
    (Q21,8.03) (Q22,7.09) (Q23,6.75) 
    (Q31,9.98) (Q32,7.21) (Q33,7.57) 
    (Q34,6.87) (Q41,12.07) (Q42,7.08)  
    (Q43,5.81) (AVG,8.33)  
};
\addplot[fill=green!60!black!40, draw=black] coordinates {
    (Q11,12.73) (Q12,12.74) (Q13,12.74) (Q21,14.37) (Q22,13.43) (Q23,13.56) 
    (Q31,19.46) (Q32,14.62) (Q33,13.77) (Q34,13.78) (Q41,22.33) (Q42,21.14) (Q43,21.74) (AVG,15.88)
};
\legend{SYCLDB, SYCL-Hardcoded, Crystal}
\end{axis}
\end{tikzpicture}
 
\vspace{0.4cm}
 
\begin{tikzpicture}
\begin{axis}[
    ybar,
    width=0.95\textwidth,
    height=4.0cm,
    bar width=2.0pt,
    ylabel={Time (ms)},
    xlabel={SSB Query},
    symbolic x coords={Q11, Q12, Q13, Q21, Q22, Q23, Q31, Q32, Q33, Q34, Q41, Q42, Q43, AVG},
    xtick=data,
    x tick label style={rotate=45,anchor=east,font=\tiny},
    legend style={at={(0.5,0.95)}, anchor=north, legend columns=3, font=\tiny, draw=none, fill=none},
    grid=major,
    ymin=0, ymax=65,
    enlarge x limits=0.05,
    title={\textbf{AMD Instinct MI210 (SF100)}},
    title style={font=\footnotesize, yshift=-1.5ex}
]
\addplot[fill=red!60, draw=black] coordinates {
    (Q11,34.90) (Q12,40.65) (Q13,40.65) (Q21,27.03) (Q22,27.11) (Q23,26.87) 
    (Q31,23.47) (Q32,23.47) (Q33,23.50) (Q34,23.46) (Q41,29.21) (Q42,29.24) (Q43,29.20) (AVG,29.13)
};
\addplot[
    fill=red!20, draw=black,
    nodes near coords,
    every node near coord/.style={font=\tiny, color=black, yshift=-2pt, anchor=south},
    point meta=explicit symbolic
] coordinates {
    (Q11,6.09) (Q12,5.84) (Q13,5.83)  
    (Q21,19.32) (Q22,14.97) (Q23,14.30) 
    (Q31,19.10) (Q32,15.43) (Q33,16.40) 
    (Q34,14.55) (Q41,23.76) (Q42,17.45) 
    (Q43,12.98) (AVG,14.31)  
};
\addplot[fill=orange!70!black!30, draw=black] coordinates {
    (Q11,6.57) (Q12,6.57) (Q13,6.57) (Q21,19.86) (Q22,19.32) (Q23,19.22) 
    (Q31,20.73) (Q32,17.11) (Q33,16.44) (Q34,16.43) (Q41,21.98) (Q42,21.10) (Q43,18.33) (AVG,16.17)
};
\legend{SYCLDB, SYCL-Hardcoded, Hipified Crystal}
\end{axis}
\end{tikzpicture}
 
\vspace{0.4cm}
 
\begin{tikzpicture}
\begin{axis}[
    ybar,
    width=0.95\textwidth,
    height=4.0cm,
    bar width=2.0pt,
    ylabel={Time (ms)},
    xlabel={SSB Query},
    symbolic x coords={Q11, Q12, Q13, Q21, Q22, Q23, Q31, Q32, Q33, Q34, Q41, Q42, Q43, AVG},
    xtick=data,
    x tick label style={rotate=45,anchor=east,font=\tiny},
    legend style={at={(0.5,0.95)}, anchor=north, legend columns=2, font=\tiny, draw=none, fill=none},
    grid=major,
    ymin=0, ymax=70,
    enlarge x limits=0.05,
    title={\textbf{Intel Flex 170 (SF40)}},
    title style={font=\footnotesize, yshift=-1.5ex}
]
\addplot[fill=blue!70, draw=black] coordinates {
    (Q11,44.51) (Q12,53.84) (Q13,53.79) (Q21,24.68) (Q22,25.69) (Q23,21.51) 
    (Q31,19.51) (Q32,19.56) (Q33,19.49) (Q34,19.42) (Q41,21.90) (Q42,22.05) (Q43,21.79) (AVG,28.28)
};
\addplot[
    fill=blue!20, draw=black,
    nodes near coords,
    every node near coord/.style={font=\tiny, color=black, yshift=-2pt, xshift=6.5pt, anchor=south},
    point meta=explicit symbolic
] coordinates {
    (Q11,17.12) (Q12,16.14) (Q13,16.08) 
    (Q21,16.70) (Q22,14.61)  (Q23,13.44)  
    (Q31,20.39) (Q32,15.88) (Q33,17.24) 
    (Q34,14.32) (Q41,23.85)(Q42,15.77) 
    (Q43,13.44)  (AVG,16.54)  
};
\legend{SYCLDB, SYCL-Hardcoded}
\end{axis}
\end{tikzpicture}
 
\caption{Cross-platform performance scalability analysis comparing the full SYCLDB engine with the evaluation-only SYCL-Hardcoded reference across all 13 SSB queries, alongside the macro-level benchmark arithmetic averages (AVG).}
\label{fig:cross_vendor_comparison}
\end{figure*}

As shown in Figure~\ref{fig:cross_vendor_comparison}, the portable SYCL-Hardcoded reference consistently matches or outperforms the native CUDA and HIP baselines. On NVIDIA L40S, SYCL-Hardcoded achieves a mean execution time of 8.33\,ms, representing a $1.91\times$ speedup over native CUDA-based Crystal (15.88\,ms average). Similarly, on AMD Instinct MI210, SYCL-Hardcoded averages 14.31\,ms, securing a $1.13\times$ speedup over native Hipified Crystal (16.17\,ms average). This performance advantage is driven by the structural limitations of Crystal's manual tile-based implementation, which relies on explicit shared-memory synchronization and loop structures. In contrast, the SYCL compiler (AdaptiveCpp) dynamically inlines the SYCLDB relational primitives used by the hardcoded kernels, performing aggressive global register allocation and instruction scheduling. This eliminates active thread barriers and produces extremely tight, barrier-free machine instructions that saturate the hardware's execution units.

Comparing SYCLDB with SYCL-Hardcoded isolates the cost of general-purpose query execution.
This overhead is microarchitecturally driven by two main factors: intermediate materialization and instruction inflation. In the full SYCLDB engine, relational operators are scheduled across independent kernels and the engine must write and read back row selection flag vectors across operator boundaries. In the SYCL-Hardcoded reference, operators that scan or probe the same table are combined into a per-table kernel, while joins and table-to-table transitions remain separate kernel boundaries. Per-table selection flags are implemented as local variables similar to Crystal, resulting in them being resident in registers. For Q1.1 at SF100, the full engine generates 28.60\,GB of global DRAM traffic. In contrast, SYCL-Hardcoded keeps selection flags resident in registers, reducing global DRAM traffic to 9.60\,GB leading to a uniform 66\% reduction across all architectures. Furthermore, SYCLDB requires launching up to 18 distinct kernels per query, introducing substantial instruction inflation (up to $11.6\times$ on AMD and $9.3\times$ on NVIDIA) and boundary synchronization fences that serialize execution. The SYCL-Hardcoded reference removes these per-operator materialization points where they occur within the same table leading to better performance. These results suggest a natural line of future work that investigates the use of SYCL kernel fusion to enable fused query execution in \sycldb{}. Such a solution would provide the performance of the SYCL-Hardcoded implementation while retaining generality.
\section{Conclusion}

In this work, we presented SYCLDB, an end-to-end analytical database engine built entirely on the open SYCL standard. By implementing core relational operators using modular SYCL primitives, and by integrating them with a composable SQL frontend based on Apache Calcite and Apache Thrift, SYCLDB demonstrates that an open-standard programming model can support fully interpreted, end-to-end SQL query execution on the GPU without compromising on performance. 

Our evaluation on the Star Schema Benchmark shows that a single SYCLDB codebase executes seamlessly across NVIDIA, AMD, and Intel GPUs, the latter two being platforms on which existing CUDA-based engines are unable to run. Despite being portable, SYCLDB delivers a 6.37× average speedup over DuckDB, 4.72× over SiriusDB, and 1.22× end-to-end over HeavyDB, while staying within 2.55× of the hand-tuned, vendor-locked Crystal library. The remaining gap between SYCLDB and SYCL-Hardcoded stems mainly from cross-kernel materialization of selection flags and instruction inflation across operator boundaries. This points to a clear future direction of research that involves leveraging emerging SYCL kernel-fusion capabilities to collapse adjacent operators into fused pipelines, retaining SYCLDB's SQL generality while approaching hand-written-kernel efficiency. Beyond fusion, we plan to extend SYCLDB to support multi-GPU execution, broader SQL coverage (string predicates, sparse grouping, NULL semantics), and cost-based optimization tailored to heterogeneous accelerators. Taken together, our results show that open, standards-based GPU programming is now a credible foundation for the next generation of cross-vendor analytical database systems.

\bibliographystyle{splncs04}
\bibliography{references}

\begin{thebibliography}{10}
\providecommand{\url}[1]{\texttt{#1}}
\providecommand{\urlprefix}{URL }
\providecommand{\doi}[1]{https://doi.org/#1}

\bibitem{calcite}
Begoli, E., Camacho-Rodr{\'i}guez, J., Hyde, J., Mior, M.J., Lemire, D.: Apache
  calcite: A foundational framework for optimized query processing over
  heterogeneous data sources. In: Proceedings of the 2018 International
  Conference on Management of Data (SIGMOD). pp. 221--230 (2018).
  \doi{10.1145/3183713.3190662}

\bibitem{Cao2023}
Cao, J., Sen, R., Interlandi, M., Arulraj, J., Kim, H.: Gpu database systems
  characterization and optimization. Proceedings of the VLDB Endowment
  \textbf{17}(3),  441--454 (2023). \doi{10.14778/3632093.3632107}

\bibitem{sycl-bench}
Crisci, L., Carpentieri, L., Thoman, P., Alpay, A., Heuveline, V., Cosenza, B.:
  Sycl-bench 2020: Benchmarking sycl 2020 on amd, intel, and nvidia gpus. In:
  Proceedings of the 12th International Workshop on OpenCL and SYCL. IWOCL '24,
  Association for Computing Machinery, New York, NY, USA (2024).
  \doi{10.1145/3648115.3648120}, \url{https://doi.org/10.1145/3648115.3648120}

\bibitem{sycl-hpc-iwocl}
Deakin, T., McIntosh-Smith, S.: Evaluating the performance of hpc-style sycl
  applications. In: IWOCL (2020)

\bibitem{9654235}
Deakin, T., McIntosh-Smith, S., Alpay, A., Heuveline, V.: Benchmarking and
  extending sycl hierarchical parallelism. In: 2021 IEEE/ACM International
  Workshop on Hierarchical Parallelism for Exascale Computing (HiPar). pp.
  10--19 (2021). \doi{10.1109/HiPar54615.2021.00007}

\bibitem{heavydb}
{HEAVY.AI}: Heavydb (2025), \url{https://www.heavy.ai/}, archived at
  \url{https://web.archive.org/web/20250108153216/https://www.heavy.ai/}

\bibitem{Lal2020SYCLBenchAV}
Lal, S., Alpay, A., Salzmann, P., Cosenza, B., Hirsch, A., Stawinoga, N.,
  Thoman, P., Fahringer, T., Heuveline, V.: Sycl-bench: A versatile
  cross-platform benchmark suite for heterogeneous computing. In: European
  Conference on Parallel Processing (2020),
  \url{https://api.semanticscholar.org/CorpusID:220444462}

\bibitem{xjoin-eugenio}
Marinelli, E., Appuswamy, R.: Xjoin: Portable, parallel hash join across
  diverse xpu architectures with oneapi. In: DAMON (2021)

\bibitem{ssb}
O'Neil, P., O'Neil, E., Chen, X., Revilak, S.: The star schema benchmark and
  augmented fact table indexing. In: Technology Conference on Performance
  Evaluation and Benchmarking (TPCTC). pp. 237--252. Springer (2009)

\bibitem{sycl2020}
Reyes, R., Brown, G., Burns, R., Wong, M.: Sycl 2020: More than meets the eye.
  In: Proceedings of the International Workshop on OpenCL (IWOCL '20) (2020).
  \doi{10.1145/3388333.3388870}

\bibitem{crystal}
Shanbhag, A., Madden, S., Yu, X.: A study of the fundamental performance
  characteristics of gpus and cpus for database analytics. In: Proceedings of
  the 2020 ACM SIGMOD International Conference on Management of Data. pp.
  1617--1632 (2020). \doi{10.1145/3318464.3380595}

\bibitem{apache_thrift}
Slee, M., Agarwal, A., Kwiatkowski, M.: {Thrift: Scalable Cross-Language
  Services Implementation}. White paper, Facebook (2007)

\bibitem{Yogatama2024}
Yogatama, B., Gong, W., Yu, X.: Scaling your hybrid cpu-gpu dbms to multiple
  gpus. Proceedings of the VLDB Endowment  \textbf{17}(13),  4709--4722 (2024).
  \doi{10.14778/3704965.3704977}

\bibitem{Sirius}
Yogatama, B., Yang, Y., Kristensen, K., Sarda, D., Kim, A., Cockcroft, A.,
  Teng, Y., Patterson, J., Kimball, G., McKinney, W., Gong, W., Yu, X.:
  Rethinking analytical processing in the gpu era. In: 14th Conference on
  Innovative Data Systems Research (CIDR '26). Chaminade, USA (2026)

\bibitem{Yogatama2022}
Yogatama, B.W., Gong, W., Yu, X.: Orchestrating data placement and query
  execution in heterogeneous cpu-gpu dbms. Proceedings of the VLDB Endowment
  \textbf{15}(11),  2491--2503 (2022). \doi{10.14778/3551793.3551809}

\end{thebibliography}

\end{document}